\RequirePackage{fix-cm}
\documentclass[smallextended]{svjour3}       % onecolumn (second format)
\smartqed  % flush right qed marks, e.g. at end of proof
\usepackage{graphicx}
\begin{document}

\title{Diffusion and memory effect in a stochastic processes and the correspondence to an information propagation in a social system}


\author{Peng Wang \and Feng-Chun Pan \and Jie Huo \and  Xu-Ming Wang }


\institute{P. Wang \and F.-C Pan \at
              School of Physics and Electronic-Electrical Engineering, Ningxia University, Yinchuan 750021, PR China \\
            %  \email{wangp_2010@163.com}           %  \\
         \and
               J. Huo \and X.-M. Wang \at
              School of Physics and Electronic-Electrical Engineering, Ningxia University, Yinchuan 750021, PR China\\
              Ningxia Key Laboratory of Intelligent Sensing for Desert Information, Yinchuan 750021, PR China\\
              Tel.: +86 13895305182\\
             \email{wang$\_$xm@126.com}
}


\maketitle

\begin{abstract}
A generalized Langevin equation is suggested to describe a system with memory($u(t,t') = \frac{1}{\Gamma (\nu )}(t - t')^\nu $) as well as with positive and negative damping. The equation can be transformed into the Fokker-Planck equation by using the Kramers-Moyal expansion. The solution of Fokker-Planck equation shows that velocity obeys a Gaussian distribution. The distribution curve will flatten as the memory parameter increases, which indicates that memory can enhance the randomness of the system. There are also some other memory effects behind this distribution, which can be characterized by calculating the transport coefficients, mean square displacement and correlation between the noise and space.  These discussions can be paralleled to a social system to understand the propagation of social ideology caused by memory.
\end{abstract}

\section{Introduction}
\label{intro}
Diffusion is one of universal phenomena in nature and therefore is one of the classical subjects of physics. It has been successfully described by continuous/discrete time random walk\cite{vot2018,cencetti2018,zhou2018,hara1979} and other models expressed by the stochastic differential
equations\cite{kemeny1986,maltba2018,sadoon2018,cerasoli2018,nyawo2018,govea2018}. As is well known, the diffusion in an open system is often influenced by stochastic fluctuation, internal and/or external force(noise). The internal and external fluctuations synchronously emerge in an open-system, and influence between themselves. So the additive noise and multiplicative noise are often considered as concomitant. 

It is noted that there are different representations of particle motion on two time scales, short-time and long-time,  in a stochastic system. The particle trajectory in the short-time is non-differentiable because of frequent collisions of it with others, so the changing state can not be described by certain differential equations, while the particle motion in the long-time can be described analytically because the non-differentiability can be eliminated or neglected in a large time scale. Therefore, the particle trajectory is non-differentiable in microscopic scale but differentiable  in macroscopic scale\cite{grigolini1999,rocco1999,metzler2001}. In other words, the randomness of a system can be ascribed to some extent to the memory of the microscopic dynamics\cite{Vacchini2016}. However, the differentiability of a  particle motion is guaranteed by the long-time scale, which may be compatible with the statistical description of a system with memory due to the fact that the statistical properties are generally embodied in macroscopic scale.

The memory effects in relevant systems have drawn much attention of physicists. Stanislavsky proved that the non-differentiable nature of the microscopic dynamics can be transmitted to the macroscopic level\cite{stanislavsky2000}. Maes investigated extension of fluctuation-response relations for non-equilibrium diffusion in stochastic systems with memory, and applied the relation to modify  the Sutherland-Einstein relation due to the strong memory\cite{maes2013}. The memory effects of anomalous diffusions are numerically simulated by Chattopadhyay\cite{chattopadhyay2009}. The recursive relationship in a random walk model\cite{diniz2017,silva2013}was transformed into the Fokker-Plank equation by Silva et al. to describe novel phenomena caused by memory\cite{silva2013}. Based on the elephant random walk model, the dependence of the diffusion on the initial condition and memory parameters is investigated\cite{hara1979,schutz2004}. 

 As is well known that the memory of individuals in social societies deeply affects the human behaviors such as opinion formation\cite{meng2018,bartolozzi2005}, cultural dynamics\cite{lizana2011}, crowd behaviors\cite{helbing1995}, etc., which are characterized by the complex outcome of many physiological and psychological processes. It is also known that physicists have made many attempts to describe the human behaviors within the framework of statistical physics based on the similarity between the components of a physical system and a social system, the  
Brownian particles and the individuals\cite{czirok1999}. As far as the specific methodology, the popular ones are master equation\cite{vazquez2008}, the Fokker-Planck equation transformed from the Langevin equation\cite{czirok1999}, as well as the extended versions of the equations expressing the dynamics on the complex networks\cite{Jedrzejewski2017,Jiang2008}. Some new conceptions such as social space, opinion space, social force, etc., are proposed to characterize the state of social particle, for instance, opinion particle\cite{Lewenstein1992,Castellano2009}. 

It seems almost inevitable that the memory of microscopic dynamics will give rise to some peculiar effects in both of physical and social systems.  Aiming at revealing the memory effect in diffusion process of a stochastic system, this paper suggests a stochastic dynamic model with memory described by a generalized Langevin equation, and investigate the spatial-temporal characteristics induced by the stochastic fluctuation or memory via solving a generalized Fokker-Planck-Kolmogorov equation derived from the generalized Langevin equation. And then, the results will be mapped to a social system to understand some of public opinions associated with the memory of individuals. 

\section{Stochastic dynamical model with memory}
\label{sec:level2}
The motion of a particle with memory is essentially a non-Markovian process, which can be described by a generalized Langevin equation as following  
\begin{equation}\label{eq:eq1}
\left\{\begin{array}{lll}
&\dot x = v & (a)\\
 &\dot v =  - \int\limits_0^t {u(t - t')\gamma (v)vdt'}- \nabla U(x) + G(v)\xi (t)&(b)
\end{array}\right.
\end{equation}
where $x$ and $v$ denote a displacement and velocity of a particle, respectively. $u(t)$ represents the memory function. For the classical ideal Markov process(without memory), the memory function is defined as $u(t - t') = \delta (t - t')$, where $\delta (t - t')$ is the Dirac function, and its the Laplace transform of the memory function can be written as $\bar u(s)=s^{-1}$. For a real system, the memory function is between the one without memory and the ideal one, the step function. So the Laplace transform of the memory function can be written as $\bar u(s)={s^\nu},  -1<\nu<0$\cite{chattopadhyay2009,schutz2004,nigmatullin1992}. Similarly, the memory function here is selected as the fractional order of which the Laplace transform is 
\begin{equation}\label{eq:eq2}
\bar u(s)=\left\{\begin{array}{lll}
&s^{\nu},  &s \ge 0\\
 &0, &s< 0
\end{array}\right.
\end{equation}
where $\gamma_{0}$ represents the damping coefficient, $\nu$ the power-law exponent of which the value ranges in $(-1,0)$.

In Eq.\ref{eq:eq1}, $\gamma(v)$ describes the non-linear damping function, which is selected as the form suggested by Rayleigh-Helmholtz, $\gamma(v)=\gamma_{0}(v^{2}-\alpha)$, where $\gamma_{0}$ represents the damping coefficient, $\alpha$ is a positive constant which divides the velocity space into two regions, the  positive damping region and the negative pumping one\cite{strutt2009,helmholtz1954}. $U(x)$ denotes a generalized potential, and is taken as $\frac{\omega^2}{2}x^2$ as well as linear gradient force $\nabla U(x)=\omega_{0}^2x$. $G(v)\xi(t)$ stands for random quantity corresponding to the Langevin force induced by fluctuation. $G(v)$ is the strength of noise and $\xi(t)$ the Gaussian white noise that is characterized by zero mean value, $\left \langle \xi(t) \right \rangle =0$. $G(v)\xi(t)$ is actually composed of the additive noise $\zeta (t)$ and multiplicative $g(v)\varsigma (t)$($g(v)$ is the strength of multiplicative noise, which can be selected as the linear function of velocity $v$). Therefore, based on the statistical relevance theory, the strength is given by $G(v)=\sqrt{2D_{0}v^{2}+4\theta\sqrt{D_{0}Q}v+2Q}$, where $\theta$ denotes the correlation strength factor between different noises, and satisfies with $0<\theta<1$. The detailed calculation please refer to Ref.\cite{wu1994}. 

The integration term in the time domain is introduced into the model because of the memory effects in the diffusion process. Therefore, the generalized Langevin equation proves to be the integro-differential one. In fact, the integration term $R(t)$ represents the time correlation function between the damping term and the memory, that is,  
\begin{equation}\label{eq:eq3}
R(t) = \int\limits_0^t {u(t - t')\gamma (v(x',t'))v(x',t')dt'} 
\end{equation}
To simplify this Eq.\ref{eq:eq1}, the Laplace transform is needed to be carried out on this function, and the inverse Laplace transform can be given by
\begin{equation}\label{eq:eq4}
\begin{array}{lll}
R(t) &={\cal{L}}^{-1}[\bar R(s)] \\
&= \frac{1}{2\pi i}\int\limits_{ - i\infty }^{ + i\infty } {\bar u(s) * \overline {\gamma (v(s))v(s)}{e^{st}}ds} \\
& = \frac{1}{\Gamma (\nu)}\int\limits_0^t {{{(t - t')}^\nu }\gamma (v(x',t'))v(x',t')dt'} \\
& = \frac{1}{\Gamma (\nu )}{t^\nu } *\left( \gamma (v(x,t))v(x,t)\right)
\end{array}
\end{equation}
where $\Gamma(\nu)$ denotes the Gamma function, ${\cal{L}}^{-1}$ the inverse Laplace transform and $\bar R(s)$ the Laplace transform is needed to be carried out on the integration term, that is, the Laplace transform of function $R(t)$.

From Eq.\ref{eq:eq4}, the memory function in a general form can be written as
\begin{equation}\label{eq:eq5}
u(t,t') = \frac{1}{\Gamma (\nu )}{(t - t')^\nu }
\end{equation}
Then Eq.\ref{eq:eq1} will be transformed into the corresponding Fokker-Planck equation
\begin{equation}\label{eq:eq6}
\frac{\partial P(x,v,t)}{\partial t}=\sum\limits_i^{a,b}{\sum\limits_{n = 1}^\infty{{{\left({-\nabla} \right)}^n}\left({D_i^{(n)}(x,v,t)P(x,v,t)}\right)}} 
\end{equation}
where $P(x,v,t)$ denotes the probability of particle with velocity $v$ at time $t$ and position $x$, $D^{(n)}_{i}(x,v,t-t')$ the Kramers-Moyal expansion coefficients, the subscript $i=a, b$ denotes the differential equations (a) and (b) in Eq.\ref{eq:eq1}, and the  superscript $n$ the order number. Defining the coefficients as the first time derivative of the central moment and divided by $n!$\cite{risken1984}, we have
\begin{equation}\label{eq:eq7}
D_i^{(n)}(x,v,t) = {\left. {\frac{1}{n!}\frac{d}{dt}{M_n}(t)} \right|_{t = 0}},i=a,b
\end{equation}
where $M_{n}(t)$ is the $n$th moment of variables. To simplify each moment, let the right terms in Eq.\ref{eq:eq1}(b) be written as
\begin{equation}\label{eq:eq8}
\left\{\begin{array}{lll}
H(x,v,t)=\frac{\gamma _0}{\Gamma(\nu )}{t^\nu }*(\gamma (v)v) + \nabla U(x)\\
\Psi (v,t)=G(v)\xi(t)
\end{array}\right.
\end{equation}
A Taylor expansion of function $H(x,v,t)$ and $\Psi(v,t)$ at $x(t')$ and $v(t')$ can be respectively given by(It is needed to point out that there are  equations $\frac{\partial H(x(t),v(t),t)}{\partial x(t)} \equiv \frac{\partial H(x(t'),v(t'),t')}{\partial x(t')}$, $\frac{\partial H(x(t),v(t),t)}{\partial v(t)} \equiv \frac{\partial H(x(t'),v(t'),t')}{\partial v(t')}$ and $\frac{\partial \Psi(v(t),t)}{\partial v(t)} \equiv \frac{\partial \Psi(v(t'),t')}{\partial v(t')}$)
\begin{equation}\label{eq:eq9}
\left\{\begin{array}{lll}
H(x(t),v(t),t)=&H(x(t'),v(t'),t')+\frac{\partial H(x(t'),v(t'),t')}{\partial x(t')}\left({x(t) - x(t')}\right)\\
 &+\frac{\partial H(x(t'),v(t'),t)}{\partial v(t')}\left({v(t)-v(t')}\right)+\cdots \\
\Psi(v(t),t)=\Psi(v(&t'),t')+\frac{{\partial \Psi(v(t'),t')}}{{\partial v(t')}}\left({v(t) - v(t')}\right)+\cdots 
\end{array}\right.
\end{equation}
To derive these Kramers-Moyal expansion coefficients, we substitute Eq.\ref{eq:eq9} into Eq.\ref{eq:eq1}(b), and obtain
\begin{equation}\label{eq:eq10}
\begin{array}{lll}
v(t + \tau )-v(t) =& \int\limits_t^{t + \tau }{H(x,v,t')dt'}+\int\limits_t^{t+\tau }{\frac{\partial H}{\partial x}\left({x(t) - x(t')}\right)dt'}+\cdots \\
&{+\int\limits_t^{t+\tau }{G(v)\xi (t')dt'}+\int\limits_t^{t+\tau }{\frac{\partial G}{\partial v}\left({v(t)-v(t')} \right)\xi (t')dt'}+ \cdots }
\end{array}
\end{equation}
where $\tau$ is the time interval.

The integration terms in Eq.\ref{eq:eq10}, $x(t)-x(t')$ and $v(t)-v(t')$, can be replaced by the first iteration of the equation itself,  which leads to 
\begin{equation}\label{eq:eq11}
\begin{array}{lll}
v(t+\tau )-v(t) =& \int\limits_t^{t + \tau } {H(x(t'),v(t'),t')dt'}+\int\limits_t^{t+\tau}{\frac{\partial H}{\partial x}\int\limits_t^{t'} {H(x,v,t'')dt''}dt'}\\
&+\int\limits_t^{t + \tau }{\frac{\partial H}{\partial v}\int\limits_t^{t'}{H(x,v,t'')dt''} dt'}+\int\limits_t^{t + \tau }{\frac{\partial H}{\partial v}\int\limits_t^{t'}{G(v)\xi (t'')dt''}dt'}+\cdots\\
&+ \int\limits_t^{t + \tau } {\Psi(v,t')\xi (t')dt'}  + \int\limits_t^{t + \tau } {\frac{\partial \Psi}{\partial v}\xi (t')\int\limits_t^{t'} {H(x,v,t'')dt''} dt'}\\
&+\int\limits_t^{t+\tau}{\frac{\partial G}{\partial v}\xi (t')\int\limits_t^{t'}{G(v)\xi (t'')dt''}dt'}+\cdots 
\end{array}
\end{equation}
Thus, the first-order and second-order moment are respectively given by
\begin{equation}\label{eq:eq12}
\begin{array}{lll}
\left\langle {v(t + \tau ) - v(t)} \right\rangle=&\int\limits_t^{t + \tau } {H(x(t'),v(t'),t')dt'}  + \int\limits_t^{t + \tau } {\frac{\partial H}{\partial x}\int\limits_t^{t'} {H(x,v,t'')dt''} dt'}\\
&+ \int\limits_t^{t + \tau } {\frac{\partial H}{\partial v}\int\limits_t^{t'} {H(x,v,t'')dt''} dt'}+\cdots \\
&{+\int\limits_t^{t+\tau } {\frac{\partial G}{\partial v}\int\limits_t^{t'}{G(v)\delta (t''-t')dt''}dt'}}+\cdots 
\end{array}
\end{equation}
and
\begin{equation}\label{eq:eq13}
\begin{array}{lll}
\left\langle {{{\left( {v(t + \tau ) - v(t)} \right)}^2}} \right\rangle =& \int\limits_t^{t+\tau }{H(x,v,t')dt'}\int\limits_t^{t+\tau }{H(x,v,t')dt'}\\  
&+ \int\limits_t^{t + \tau } {\frac{\partial H}{\partial x}\int\limits_t^{t'} {H(x,v,t'')dt''} dt'}+\cdots\\
&{ + \int\limits_t^{t + \tau } {G(v)\int\limits_t^{t'} {G(v)\delta (t''-t')dt''} dt'}}+\cdots
\end{array}
\end{equation}
Because of $\int\limits_t^{t'}{G(v(t''))\delta (t''-t')dt''}=G(v(t'))$, the Kramers-Moyal expansion coefficients can be written as the following when $\tau \to 0$
\begin{equation}\label{eq:eq14}
\begin{array}{lll}
D_b^{(1)}(x,v,t) &= \frac{1}{1!}\mathop {\lim }\limits_{\tau  \to 0} \frac{1}{\tau }\left\langle {v (t + \tau )-v(t)} \right\rangle \\
& = H(x,v,t) + \frac{\partial G(v)}{\partial v}G(v)
\end{array}
\end{equation}
\begin{equation}\label{eq:eq15}
\begin{array}{lll}
D_b^{(2)}(x,v,t)&=\frac{1}{\rm{2}!}\mathop{\lim}\limits_{\tau  \to 0}\frac{1}{\tau }\left\langle{v(t+\tau )-v(t)}\right\rangle \\
&= {\left( {G(v)} \right)^2}
\end{array}
\end{equation}
\begin{equation}\label{eq:eq16}
D_b^{(n)}(x,v,t) = 0, for, n \ge 3
\end{equation}
Similarly, the Kramers-Moyal expansion coefficients, corresponding to Eq.\ref{eq:eq1}(a), can be written as
\begin{equation}\label{eq:eq17}
\left\{\begin{array}{lll}
D_a^{(1)}(x,v,t) = v\\
D_a^{(n)}(x,v,t) = 0, for, n \ge 2
\end{array}\right.
\end{equation}
So we obtain the Fokker-Planck equation 
\begin{equation}\label{eq:eq18}
\begin{array}{lll}
\frac{\partial P(x,v,t)}{\partial t} = &- \frac{\partial }{\partial v}D_b^{(1)}(x,v,t)P(x,v,t) + \frac{1}{2}\frac{{{\partial ^2}}}{{\partial {v^2}}}D_b^{(2)}(x,v,t)P(x,v,t) \\
&- \frac{\partial }{\partial x}D_a^{(1)}(x,v,t)P(x,v,t)
\end{array}
\end{equation}
Substituting Eqs.\ref{eq:eq14}-\ref{eq:eq17} into Eq.\ref{eq:eq18}, it becomes
%\begin{widetext}
\begin{equation}\label{eq:eq19}
\begin{array}{lll}
\frac{\partial P(x,v,t)}{\partial t} =& - \frac{\partial }{\partial v}\left( {\frac{\gamma _0}{\Gamma (\nu )}{t^\nu } * \gamma (v)v + \nabla U(x) + \frac{{\partial G(v)}}{{\partial v}}G(v)} \right)P\\
&+ \frac{1}{2}\frac{\partial ^2}{\partial {v^2}}{\left( {G(v)} \right)^2}P - v\frac{\partial P}{\partial x}
\end{array}
\end{equation}
Here, we introduce the transition probability density $P(\bf{\sigma },t + \tau \left| {\bf{\sigma '},t} \right.)$, which can correlate the probability density at time $t+\tau$ with the one at $t$. The correlation is given by the classical Chapman-Kolmogorov equation
\begin{equation}\label{eq:eq20}
P(\bf{\sigma },t+\tau) = \int {P(\bf{\sigma},t+\tau \left| {\bf{\sigma'},t} \right.)P(\bf{\sigma'},t)d\bf{\sigma'}} 
\end{equation}
where $\bf{\sigma}=(x,v)$ denotes a generalized vector space, which is composed of the generalized displacement and the velocity at time $t+\tau$. 
Similarly, $\bf{\sigma'}=(x',v')$ denotes a vector space at time $t$. 

Eq.\ref{eq:eq20} becomes
\begin{equation}\label{eq:eq21}
P(\bf{\sigma },t + \tau \left| {\bf{\sigma '},t} \right.) = \int {\delta (\bf{y} - \bf{\sigma })P(\bf{y},t + \tau \left| {\bf{\sigma '},t} \right.)d\bf{y}} 
\end{equation}
by the identity with the delta function 
\begin{equation}\label{eq:eq22}
\begin{array}{lll}
\delta (\bf{y} - \bf{\sigma }) &= \delta (\bf{\sigma '} - \bf{\sigma } + \bf{y} - \bf{\sigma '})\\
&= \sum\limits_{n = 0}^\infty  {\frac{{{{\left( {\bf{y} - \bf{\sigma '}} \right)}^n}}}{n!}} {\left({ - \frac{\partial }{{\partial \bf{\sigma }}}} \right)^n}\delta (\bf{\sigma '} - \bf{\sigma })
\end{array}
\end{equation}
And then, Eq.\ref{eq:eq21} changes to
\begin{equation}\label{eq:eq23}
\begin{array}{lll}
P(\bf{\sigma },t + \tau \left| {\bf{\sigma '},t} \right.) &= \sum\limits_{n = 0}^\infty  {\frac{1}{{n!}}} {\left( { - \frac{\partial }{{\partial \bf{\sigma }}}} \right)^n}\int {{{\left( {\bf{y} - \bf{\sigma '}} \right)}^n}P(\bf{y},t + \tau \left| {\bf{\sigma '},t} \right.)dy} \delta (\bf{\sigma '} - \bf{\sigma })\\
 &= \left({1+ \sum\limits_{n = 0}^\infty  {\sum\limits_i^{a,b} {\frac{1}{{n!}}{{\left( { - \frac{\partial }{{\partial \bf{\sigma }}}} \right)}^n}M_i^n(t)} } } \right)\delta(\bf{\sigma}-\bf{\sigma'})
\end{array}
\end{equation}
%\end{widetext}
Obviously, the Kramers-Moyal expansion coefficients are given by
\begin{equation}\label{eq:eq24}
\frac{M_i^n(t)}{n!} = D_i^{(n)}(\bf{\sigma },t)\tau  + O({\tau ^2}),i=a,b
\end{equation}
By substituting Eq.\ref{eq:eq24} into Eq.\ref{eq:eq23}, we have
%\begin{widetext}
\begin{equation}\label{eq:eq25}
\begin{array}{lll}
P(\bf{\sigma },t + \tau \left|{\bf{\sigma '},t} \right.) =& \left( {1 - \frac{\partial }{\partial v}\left( {\frac{\gamma _0}{\Gamma (\nu)}{t^\nu } * \gamma (v)v + \nabla U(x) + \frac{\partial G(v)}{\partial v}G(v)} \right)\tau } \right.\\
&\left.{+\frac{1}{2}\frac{\partial ^2}{\partial {v^2}}{{\left( {G(v)} \right)}^2}\tau  - \left( {v\frac{\partial }{\partial x}} \right)\tau  + O({\tau ^2})} \right)\delta(\bf{\sigma}-\bf{\sigma'})
\end{array}
\end{equation}
To solve Eq.\ref{eq:eq25}, we conduct the following simplifications: neglecting the term of 2-order in $\tau$ and replacing $\delta$ function $\delta(\bf{\sigma}-\bf{\sigma'})=\delta(x-x')\delta(v-v')$ by their Fourier transforms  
\begin{equation}\label{eq:eq26}
\begin{array}{lll}
P(\bf{\sigma },t + \tau \left| {\bf{\sigma '},t} \right.) =& \frac{1}{4\pi ^2}\exp \left[ { - \frac{\partial }{\partial v}\left( {\frac{\gamma _0}{\Gamma (\nu )}{t^\nu } * \gamma (v)v + \nabla U(x) + \frac{\partial G(v)}{\partial v}G(v)} \right)\tau } \right.\\
&{\left. { + \frac{1}{2}\frac{\partial ^2}{\partial {v^2}}{{\left( {G(v)} \right)}^2}\tau  - \left( {v\frac{\partial }{\partial x}} \right)\tau } \right]\int\limits_{ - \infty }^{ + \infty } {{e^{ik_{2} (x - x')}}dk_{2} } \int\limits_{-\infty}^{+\infty} {{e^{ik_{1}(v - v')}}dk_{1}} }
\end{array}
\end{equation}
The differential operator of velocity and that of displacement in this equation can be replaced by their integral transforms, so the equation is solved, and we have
\begin{equation}\label{eq:eq27}
\begin{array}{lll}
P(x,v,t + \tau \left| {x',v',t} \right.)&= \frac{1}{4\pi ^2}\int\limits_{- \infty }^{ + \infty } {\exp \left[{ - K(x,v,t)i{k_1}\tau }{ - \frac{1}{2}k_1^2\tau {{\left( G(v) \right)}^2} + i{k_1}(v - v')} \right]d{k_1}}\\
&{\times\int\limits_{ - \infty }^{ + \infty } {\exp \left[ {i{k_2}( - v\tau + x - x')} \right]d{k_2}} }\\
&=\frac{1}{\sqrt {8{\pi ^3}\tau {\left[ G(v) \right)}^2}}{\exp{\left[{-\frac{{\left({v - v' -\tau K(x,v,t)}\right)}^2}{4\tau {{\left( {G(v)} \right)}^2}}} \right]}}
\end{array}
\end{equation}
where $K(x,v,t)={\frac{\gamma _0}{\Gamma (\nu )}{t^\nu } * \gamma (v)v + \nabla U(x) + \frac{\partial G(v)}{\partial v}G(v)}$. The stable solution of Eq.\ref{eq:eq19}, obtained as $t$ trends to infinity, is in the form
\begin{equation}\label{eq:eq28}
P(x,v,\infty ) = \exp \left[ - \frac{\omega _0^2}{2}{x^2} - \frac{v^2}{2} \right]
\end{equation}
Therefore, the probability density can be written as
\begin{equation}\label{eq:eq29}
P(x,v,t;x',v',\infty) = P(x,v,t|x',v',\infty)P(x',v',\infty)
\end{equation}

\section{The diffusion characteristic of particles}
\label{sec:level2}
Now lets present the discussions on the transport characteristics based on the solution, Eq. \ref{eq:eq29}. The parameters are selected as ${\omega _0} = 5.0 \times {10^{ - 2}},{r_0}=1.0 \times {10^{ - 2}},\theta = 0.9,\alpha=0.3,\nu=-0.2$. Based on the expression of probability density, we investigate the transport characteristics via calculating relevant statistical quantities, such as diffusion coefficient, viscosity coefficient and thermal conductivity coefficient.

\subsection{The characteristics of the velocity distribution}
\label{sec:1}
\begin{figure}[htbp!]
   \centering
    \includegraphics[width=1.0\textwidth]{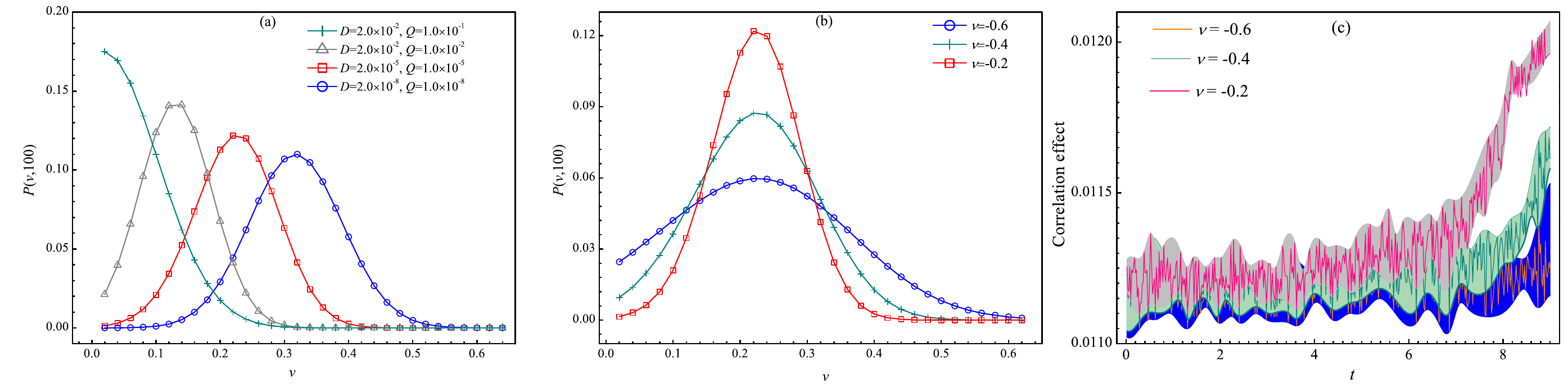}
    \caption{The velocity distribution under different parameter combinations. (a) The Gaussian distribution, the steady distribution of velocity flattens with the decrease of the correlation intensity denoted by $D$ and $Q$; (b) The steady distribution flattens with the characteristic parameter $\nu$ as $D, Q$ is $2.0\times 10^{-5}$, $1.0\times10^{-5}$, respectively; (c) The dependence of the correlation between the noise and space on time with different parameters of memory.}
\label{fig:fig1}
\end{figure}

Fig.~\ref{fig:fig1} shows that the Gaussian distribution dominating the steady state. The results are different from Refs.\cite{han2005,han200592} and our previous work without memory\cite{han2005,han200592} where there is non-equilibrium spike, appearing on the background of such steady distribution, caused by the correlation between noise and space. Fig.~\ref{fig:fig1}(a) shows, the peak of the Gaussian distribution moves in the direction of velocity increasing and transforms into the flatting pattern with decrease of correlation intensity of noise. The shift of the peak indicates that the expected value of velocity and the standard deviation will increase. The flattening of the distribution curve implies that the decrease of the correlation of noise will homogenize the velocity distribution to weaken the correlation between particles. The reason may be that the memory maintains the state of every particle and restrains the correlation among them. It is also proved by Fig.~\ref{fig:fig1}(b), where the probability density gradually flattens with increase of the characteristic parameter of memory, $\nu$(of which the value $-1$ and $0$ respectively corresponds to the ideal memory and the absence of memory). The results indicate that the second moment of velocity increases, which also implies that the randomness of the particle is strengthened. Compared with the situation without memory, memory can restrain the correlation between the noise and space. This conclusion can be obtained by extrapolating from the results shown by  Fig.~\ref{fig:fig1}(c) that correlation between the noised and space decreases with the increase of the parameter of memory. Meanwhile, the difference between dissipative region and pumping region disappears. This implies that the energy captured from environment in the pumping region decreased, and stochastic collision in the dissipative region is weakened by the memory effects. In fact, it has been stressed above that the memory can enhance the ability of the system keeping its state but weaken the response of the system to the stochastic fluctuation.

\subsection{The characteristics of the diffusion}
\label{sec:2}
As is well known, if a macroscopic system is at a non-equilibrium state, a transport process will often occur to establish a new equilibrium. Generally, the transport process is irreversible, and the establishment of the new equilibrium will take some limited time, named relaxation time. The characteristics of the process can be described by the transport coefficients such as diffusion, viscosity and thermal conductivity coefficient. Based on the statistical correlation of Green-Kubo\cite{kubo1991,mahan2000,deo1966,lixb2010,pobert1967}, these coefficients can be given by 
\begin{equation}\label{eq:eq30}
\cases{
D_{\rm{eff}} = \frac{1}{2dt}\left\langle {\left( x(t) - \left\langle x \right\rangle  \right)}^2 \right\rangle \\
\eta  = \frac{1}{2dLt}\left\langle {\left( \rho vx - \left\langle {\rho vx} \right\rangle  \right)}^2 \right\rangle \\
\kappa = \frac{1}{2d{T^2}Lt}\left\langle {\left( xe - \left\langle {xe} \right\rangle \right)}^2 \right\rangle 
}
\end{equation}
where $D_{\rm{eff}},\eta,\kappa$ denote diffusion, viscosity and thermal conductivity coefficient, respectively. $L$ represents the spatial scale, $d$ the spatial dimension, which is selected as $d=1$ in the model, $e$ the statistical kinetic energy and $T$ the thermodynamic temperature. The thermodynamic temperature is classically defined as
\begin{equation}\label{eq:eq31}
T = k_B^{-1}\left\langle \rho{v^2} \right\rangle 
\end{equation}
where $k_{B}$ is the Boltzmann constant.

\begin{figure}[htbp!]
\centering
   \includegraphics[width=0.9\textwidth]{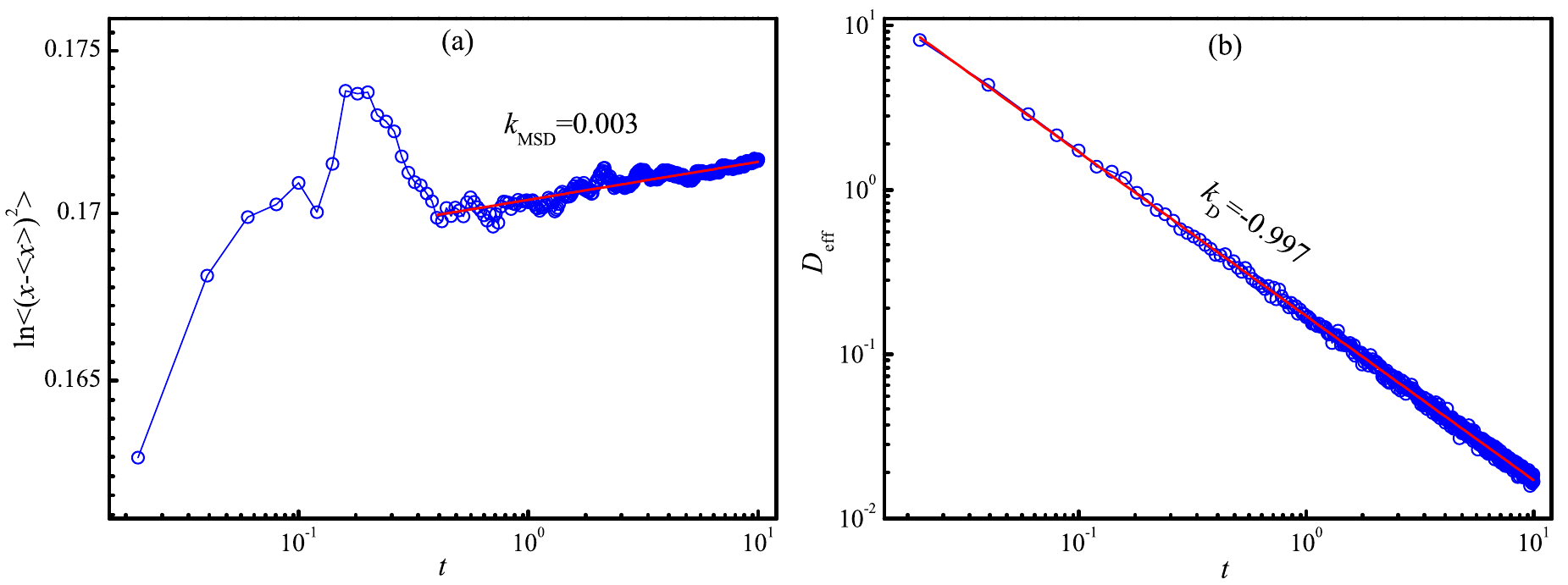}
\caption{The scaling relations of (a) the mean square displacement and (b) the diffusion coefficient relate to time. }
\label{fig:fig2}
\end{figure}

Fig.~\ref{fig:fig2} shows the dependence of the mean square displacement and diffusion coefficient on time. Fig.~\ref{fig:fig2}(a) demonstrates that the mean square displacement depends on time partly in the power law with exponent $k_{\rm{MSD}} = 0.003$. Obviously, the exponent is very closed to $0$, and therefore the mean square displacement tends to a constant($\left\langle {(x-\left\langle {x} \right\rangle)^{2}} \right\rangle \approx 0.17$). Fig.~\ref{fig:fig2}(b) shows that the diffusion is also related to time in power law with exponent $k_{\rm{D}}=-0.997$ which is very closed to $-1$. The exponents for both of the mean square displacement and diffusion coefficient imply that the diffusion is a sub-diffusion\cite{cairoli2015,selmeczi2008,selmeczi2005,bronstein2009,caspi2000,bruno2009,greenenko2004,harris2012}.  Compared this diffusion process with that without memory\cite{wang2018,wangjh2017,lewandowski2013}, it is obvious that the non-equilibrium behaviors induced by correlation between the noise and the space do not appear. These results support the conclusion above that the memory restrains the randomness of the system.

\begin{figure}[htbp!]
\centering
   \includegraphics[width=0.45\textwidth]{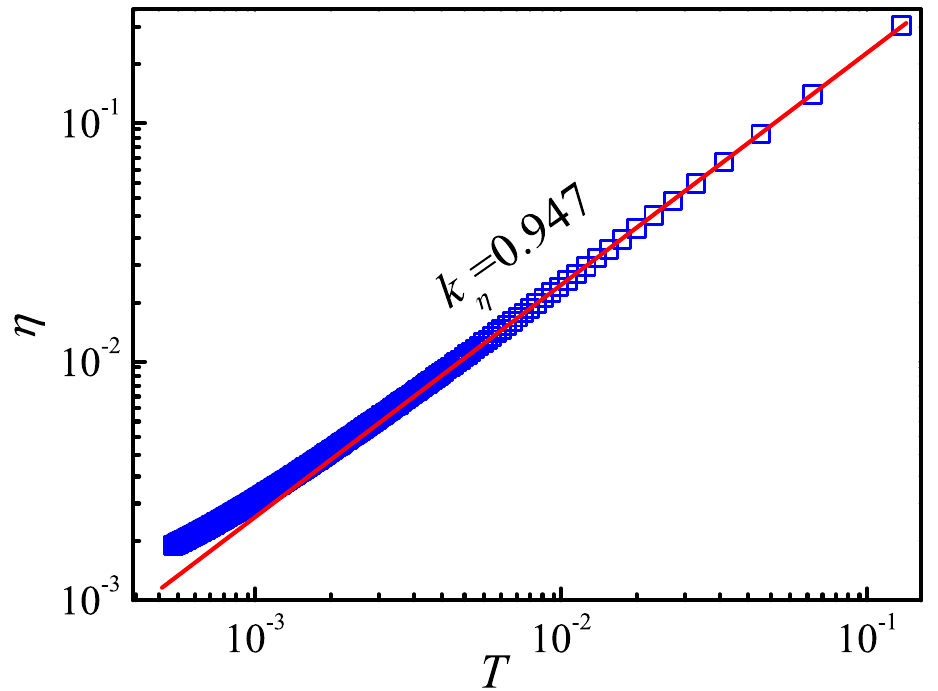}
\caption{The scaling law between the viscosity coefficient and statistical temperature.}
\label{fig:fig3}
\end{figure}

As shown in Fig.~\ref{fig:fig3}, the dependence of the viscosity coefficient on statistical temperature obeys a power law with a plus exponent $k_{\eta}=0.947$. This result might be consistent with the property of a sparse particle system in which the viscosity mainly originates from the collision among the particles. With increase of temperature, the collision frequency and the velocity increase, so the exchange of mass and momentum among the particles strengthens, which leads to the increase of the viscosity coefficient. In fact, the similar results are observed in the sparse systems such as Fermi gas\cite{pearson1994,bruun2005,bluhm2017}, ideal fluid\cite{joseph2015,wlazlowski2015,mauro2009,fomin2012,dymond1981}, etc.

\begin{figure}[htbp!]
\centering
   \includegraphics[width=0.45\textwidth]{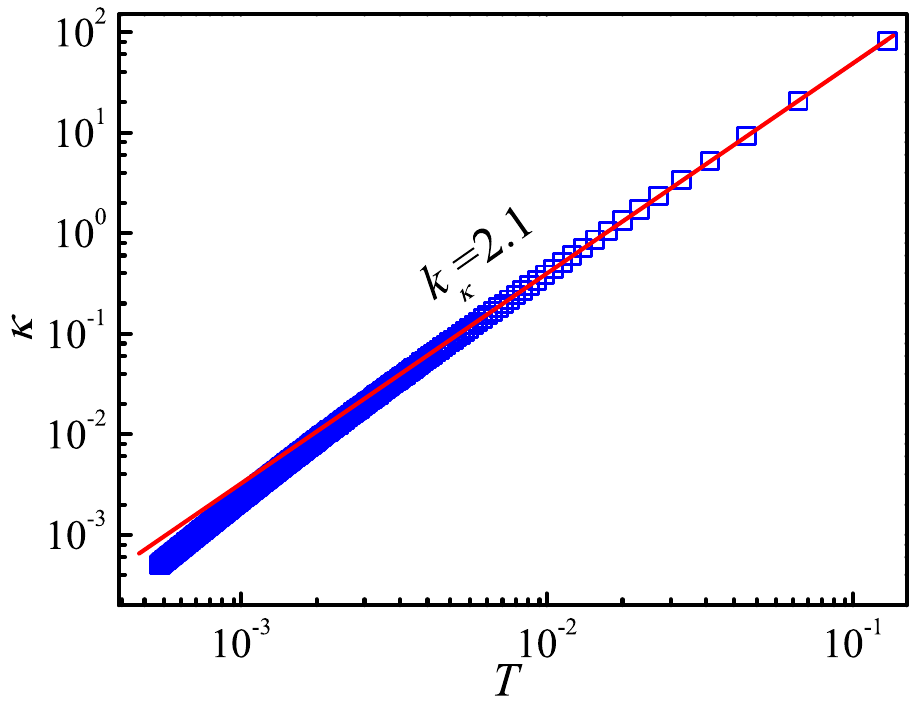}
\caption{Scaling law between the thermal conductivity coefficient and temperature.}
\label{fig:fig4}
\end{figure}

Fig.~\ref{fig:fig4} shows the dependence of the thermal conductivity on temperature is dominated by the power law with scaling exponent $k_{\kappa}=2.1$. The similar relations are observed in some systems such as amorphous solid, ideal fluid and sparse clathrate hydrate\cite{krivchikov2005,krivchikov2006,alshaikhi2010}. Especially, the simulation results in a one-dimensional molecular chain indicate that the thermal conductivity increases monotonously with temperature as the chain density is less than $1$\cite{savin2014}. So we can conclude that there is the same mechanism, collision/interaction among the particle, governing the relation between the viscosity and temperature and that between the thermal conductivity and temperature. 
  
\section{The correspondence of the diffusion characteristics to social behaviors }
\label{sec:level3}
A social system can be regarded as a classical open system composed of the so-called quasi-particles describing public opinion, ideology, etc. It is evident that there is some similarities between the physical systems and the social systems, as well as that between the corresponding units, for instance, the particles and opinion particles\cite{romensky2014,martins2015}. So the regularities dominating the behaviors of a physical system can be paralleled to a corresponding social system to interpret or understand the social behaviors. The  change of a sociology is generally deeply influenced by its historical state, then the discussions above on the diffusion behaviors of the system with memory can be naturally used to describe the propagation of the opinion, idea, belief, etc. 

The prerequisite for doing so is that we need to re-define the variables and parameters of the physical system so as to describe the social system reasonably.  The variables in Eq.\ref{eq:eq1} might become the "generalized" ones, and the corresponding new definitions are suggested in our previous work(which will be reported elsewhere) where we proposed an information space in which the position of an opinion particle relative to the origin is used to define the generalized displacement. In other words, the generalized displacement is defined as the quantity of information carried by a quasi-particle in a social system. The velocity, the changing rate of the generalized displacement/the quantity of information, is defined as the sensitivity of the quasi-particle to an event. The effect of noise is that caused by the influence of the inter and/or outer environment that is introduced into the system via the coupling of it with the generalized displacement. 

Now we might understand the information propagation in the process such as disseminations of opinion, idea, belief, etc. by means of the characteristic coefficients describing the diffusion process with memory. The overall characteristics of the system is depicted by Fig.~\ref{fig:fig1}. As shown by Fig.~\ref{fig:fig1}(a), the mean velocity of information propagation increases when the random force, originating from the social environment, weakens. As shown by Fig.~\ref{fig:fig1}(b), the memory enhances the randomness via decreasing the coupling between the individuals by comparing with that the memory strengthens the independence of the individuals. The scaling exponent, $k_{\rm{D}}\to -1$, for the power law governing the variation of the diffusion coefficient with time as shown in Fig.~\ref{fig:fig2} indicates that the information propagation process is an extreme sub-diffusion. The corresponding dynamic mechanism can be interpreted as that the information propagation is achieved mainly via colliding between the quasi-particles due to the fact presented by Fig.~\ref{fig:fig3} and Fig.~\ref{fig:fig4}. The two figures show that the viscosity coefficient and thermal conductivity coefficient respectively depend on temperature in power law manner. The plus scaling exponents indicate that the information propagate mainly by means of colliding between quasi-particles. And then, one may draw a conclusion that the memory can enhance the individualism and the randomness of system(the distribution curve flattening), and therefore strengthen the stability for the structure of the social group. It should be stressed that the main propagation mechanism is the collision between quasi-particles.

\section{Conclusion and Discussion}
\label{sec:level4}
In this article, the transport behaviors in the open particle system with memory are investigated by the Fokker-Planck equation transformed from a generalized Langevin equation. The solution brings to light the details of the transport behaviors, which can be paralleled to the social system formed by quasi-particles to describe the propagation of idea, opinion, belief, etc..

In the particle system, the curve of the steady distribution of velocity moves in the direction of velocity increasing and transforms into the flatting pattern with decrease of correlation intensity of noise, and also gradually flattens with increase of the memory. These imply that the memory can enhance the ability of the system keeping its state but weaken the response of the system to the stochastic fluctuation. The minus scaling exponent for the power law governing the variation of the diffusion coefficient with time indicates that the process is a sub-diffusion. The plus scaling exponents for the power laws governing the relation between the viscosity coefficient and temperature as well as that between the thermal conductivity coefficient and temperature reveal that the corresponding transport mechanism can be mainly attributed to the collision between particles. The similar relations are observed in some systems such as sparse clathrate hydrate\cite{krivchikov2005,krivchikov2006,andersson1996} and single crystal\cite{alshaikhi2010,yang2010}. 

Based on the similarity between the diffusion in physical system and the propagation of idea, opinion, belief, etc., in social system, the results and discussions in the former are paralleled to the latter. The mean velocity of information propagation increases when the random force, random disturbance originating from the social environment, weakens. The memory enhances the randomness via decreasing the coupling between the individuals. In other words, the memory increases the independence of the individuals, the collision, interaction, among them weakens, and therefore the transport process is sub-diffusion. Generally speaking, the memory can enhance the structure stability of a social group. It may be a important inspiration for us.

\begin{acknowledgements}
This study is supported by National Natural Science Foundation of China under Grant Nos. 11665018.
\end{acknowledgements}

%\begin{acknowledgements}
%\end{acknowledgements}


\providecommand{\newblock}{}
\begin{thebibliography}{10}
\expandafter\ifx\csname url\endcsname\relax
  \def\url#1{{\tt #1}}\fi
\expandafter\ifx\csname urlprefix\endcsname\relax\def\urlprefix{URL }\fi
\providecommand{\eprint}[2][]{\url{#2}}

\bibitem{vot2018}
Vot F~L and Yuste S~B 2018 {\em Physical Review E\/} {\bf 98} 042117

\bibitem{cencetti2018}
Cencetti G, Battiston F, Fanelli D and Latora V 2018 {\em Physical Review E\/}
  {\bf 98} 052302

\bibitem{zhou2018}
Zhou Z~Z, Grimm J, Fang S, Deng Y and Garoni T~M 2018 {\em Physical Review
  Letters\/} {\bf 121} 185701

\bibitem{hara1979}
Hara H 1979 {\em Physical Review B\/} {\bf 20} 4062--4068

\bibitem{kemeny1986}
Kemeny G, Mahanti S~D and Kaplan T~A 1986 {\em Physical Review B\/} {\bf 34}
  6288--6294

\bibitem{maltba2018}
Maltba T, Gremaud P~A and Tartakovsky D~M 2018 {\em Journal of Computational
  Physics\/} {\bf 367} 87--101

\bibitem{sadoon2018}
Sadoon A~A and Wang Y 2018 {\em Physical Review E\/} {\bf 98} 042411

\bibitem{cerasoli2018}
Cerasoli S, Dotsenko V, Oshanin G and Rondoni L 2018 {\em Physical Review E\/}
  {\bf 98} 042149

\bibitem{nyawo2018}
Nyawo P~T and Touchette H 2018 {\em Physical Review E\/} {\bf 98} 052103

\bibitem{govea2018}
Olais-Govea J~M, L{\'{o}}pez-Flores L, Ch{\'{a}}vez-P{\'{a}}ez M and
  Medina-Noyola M 2018 {\em Physical Review E\/} {\bf 98} 040601

\bibitem{grigolini1999}
Grigolini P, Rocco A and West B~J 1999 {\em Physical Review E\/} {\bf 59}
  2603--2613

\bibitem{rocco1999}
Rocco A and West B~J 1999 {\em Physica A Statistical Mechanics \& Its
  Applications\/} {\bf 265} 535--546

\bibitem{metzler2001}
Metzler R 2001 {\em Fractals\/} {\bf 09} 373--374

\bibitem{Vacchini2016}
Vacchini B 2016 {\em Physical Review Letters\/} {\bf 117}

\bibitem{stanislavsky2000}
Stanislavsky A~A 2000 {\em Physical Review E\/} {\bf 61}(5) 4752--4759

\bibitem{maes2013}
Maes C, Safaverdi S, Visco P and van Wijland F 2013 {\em Physical Review E\/}
  {\bf 87} 022125

\bibitem{chattopadhyay2009}
Chattopadhyay A~K 2009 {\em Physical Review E\/} {\bf 80} 011144

\bibitem{diniz2017}
Diniz R~M~B, Cressoni J~C, da~Silva M~A~A, Mariz A~M and de~Ara{\'{u}}jo J~M
  2017 {\em Physical Review E\/} {\bf 96} 062143

\bibitem{silva2013}
da~Silva M~A~A, Cressoni J~C, Sch{\"u}tz G~M, Viswanathan G~M and Trimper S
  2013 {\em Physical Review E\/} {\bf 88} 022115

\bibitem{schutz2004}
Sch{\"u}tz G~M and Trimper S 2004 {\em Physical Review E\/} {\bf 70} 045101(R)

\bibitem{meng2018}
Meng X~F, Van~Gorder R~A and Porter M~A 2018 {\em Phys. Rev. E\/} {\bf 97}(2)
  022312

\bibitem{bartolozzi2005}
Bartolozzi M, Leinweber D~B and Thomas A~W 2005 {\em Phys. Rev. E\/} {\bf
  72}(4) 046113

\bibitem{lizana2011}
Lizana L, Mitarai N, Sneppen K and Nakanishi H 2011 {\em Phys. Rev. E\/} {\bf
  83}(6) 066116

\bibitem{helbing1995}
Helbing D and Moln\'ar P 1995 {\em Phys. Rev. E\/} {\bf 51}(5) 4282--4286

\bibitem{czirok1999}
Czir\'ok A, Barab\'asi A~L and Vicsek T 1999 {\em Phys. Rev. Lett.\/} {\bf
  82}(1) 209--212

\bibitem{vazquez2008}
Vazquez F, Egu\'{\i}luz V~M and Miguel M~S 2008 {\em Phys. Rev. Lett.\/} {\bf
  100}(10) 108702

\bibitem{Jedrzejewski2017}
J{\k{e}}drzejewski A 2017 {\em Physical Review E\/} {\bf 95}

\bibitem{Jiang2008}
Jiang L~L, Hua D~Y, Zhu J~F, Wang B~H and Zhou T 2008 {\em The European
  Physical Journal B\/} {\bf 65} 251--255

\bibitem{Lewenstein1992}
Lewenstein M, Nowak A and Latan{\'{e}} B 1992 {\em Physical Review A\/} {\bf
  45} 763--776

\bibitem{Castellano2009}
Castellano C, Fortunato S and Loreto V 2009 {\em Reviews of Modern Physics\/}
  {\bf 81} 591--646

\bibitem{nigmatullin1992}
Nigmatullin R~R 1992 {\em Theoretical and Mathematical Physics\/} {\bf 90}
  242--251 ISSN 1573-9333

\bibitem{strutt2009}
Strutt J~W 2009 {\em The Theory of Sound\/} (Cambridge University Press)

\bibitem{helmholtz1954}
Helmholtz H 1954 {\em On the Sensations of Tone\/} (Doubleday,New York)

\bibitem{wu1994}
Cao L, Wu D~J and Ke S~Z 1995 {\em Physical Review E\/} {\bf 52} 3228--3231

\bibitem{risken1984}
Risken H 1984 {\em The Fokker-Planck Equation\/} (Springer Berlin Heidelberg)

\bibitem{han2005}
Han Y~X, Li J~H and Chen S~G 2005 {\em Communications in Theoretical Physics\/}
  {\bf 44} 226--230

\bibitem{han200592}
Han Y~X, Li J~H, Zhao Y~K and Chen S~G 2005 {\em Communications in Theoretical
  Physics\/} {\bf 43} 92--96

\bibitem{kubo1991}
Kubo R, Toda M and Hashitsume N 1991 {\em Statistical Physics {II}\/} (Springer
  Berlin Heidelberg)

\bibitem{mahan2000}
Mahan G~D 2000 {\em Many-Particle Physics\/} (Springer {US})

\bibitem{deo1966}
Deo B and Behera S~N 1966 {\em Physical Review\/} {\bf 141} 738--741

\bibitem{lixb2010}
Li X, Maute K, Dunn M~L and Yang R 2010 {\em Physical Review B\/} {\bf 81}
  245318

\bibitem{pobert1967}
PETERSON R~L 1967 {\em Reviews of Modern Physics\/} {\bf 39} 69--77

\bibitem{cairoli2015}
Cairoli A and Baule A 2015 {\em Physical Review Letters\/} {\bf 115} 110601

\bibitem{selmeczi2008}
Selmeczi D, Li L, Pedersen L~I, Nrrelykke S~F, Hagedorn P~H, Mosler S, Larsen
  N~B, Cox E~C and Flyvbjerg H 2008 {\em The European Physical Journal Special
  Topics\/} {\bf 157} 1--15

\bibitem{selmeczi2005}
Selmeczi D, Mosler S, Hagedorn P~H, Larsen N~B and Flyvbjerg H 2005 {\em
  Biophysical Journal\/} {\bf 89} 912--931

\bibitem{bronstein2009}
Bronstein I, Israel Y, Kepten E, Mai S, Shav-Tal Y, Barkai E and Garini Y 2009
  {\em Physical Review Letters\/} {\bf 103} 018102

\bibitem{caspi2000}
Caspi A, Granek R and Elbaum M 2000 {\em Physical Review Letters\/} {\bf 85}
  5655--5658

\bibitem{bruno2009}
Bruno L, Levi V, Brunstein M and Desp{\'{o}}sito M~A 2009 {\em Physical Review
  E\/} {\bf 80} 011912

\bibitem{greenenko2004}
Greenenko A, Chechkin A and Shul{\textquotesingle}ga N 2004 {\em Physics
  Letters A\/} {\bf 324} 82--85

\bibitem{harris2012}
Harris T~H, Banigan E~J, Christian D~A, Konradt C, Wojno E~D~T, Norose K,
  Wilson E~H, John B, Weninger W, Luster A~D, Liu A~J and Hunter C~A 2012 {\em
  Nature\/} {\bf 486} 545--548

\bibitem{wang2018}
Wang J~H, Li H~Y, Chen Y~P, Liu S~Y, Yan P, Shen Y, Guo J~S and Fang F 2018
  {\em Environmental Science and Pollution Research\/} {\bf 25} 9797--9805

\bibitem{wangjh2017}
Wang J~H, Chen Y~P, Dong Y, Wang X~X, Guo J~S, Shen Y, Yan P, Ma T~F, Sun X~Q,
  Fang F and Wang J 2017 {\em Environmental Pollution\/} {\bf 229} 199--209

\bibitem{lewandowski2013}
Lewandowski Z and Beyenal H 2013 {\em Fundamentals of Biofilm Research, Second
  Edition\/} ({CRC} Press)

\bibitem{pearson1994}
Pearson D~S, Fetters L~J, Graessley W~W, Strate G~V and von Meerwall E 1994
  {\em Macromolecules\/} {\bf 27} 711--719

\bibitem{bruun2005}
Bruun G~M and Smith H 2005 {\em Physical Review A\/} {\bf 72} 043605

\bibitem{bluhm2017}
Bluhm M, Hou J and Sch{\"a}fer T 2017 {\em Physical Review Letters\/} {\bf 119}
  065302

\bibitem{joseph2015}
Joseph J, Elliott E and Thomas J 2015 {\em Physical Review Letters\/} {\bf 115}
  020401

\bibitem{wlazlowski2015}
Wlaz{\l}owski G, Quan W and Bulgac A 2015 {\em Physical Review A\/} {\bf 92}
  063628

\bibitem{mauro2009}
Mauro J~C, Allan D~C and Potuzak M 2009 {\em Physical Review B\/} {\bf 80}
  094204

\bibitem{fomin2012}
Fomin Y~D, Brazhkin V~V and Ryzhov V~N 2012 {\em Physical Review E\/} {\bf 86}
  011503

\bibitem{dymond1981}
Dymond J~H and Young K~J 1981 {\em International Journal of Thermophysics\/}
  {\bf 2} 237--247

\bibitem{krivchikov2005}
Krivchikov A~I, Manzhelii V~G, Korolyuk O~A, Gorodilov B~Y and Romantsova O~O
  2005 {\em Physical Chemistry Chemical Physics\/} {\bf 7} 728--730

\bibitem{krivchikov2006}
Krivchikov A~I, Gorodilov B~Y, Korolyuk O~A, Manzhelii V~G, Romantsova O~O,
  Conrad H, Press W, Tse J~S and Klug D~D 2006 {\em Physical Review B\/} {\bf
  73}

\bibitem{alshaikhi2010}
AlShaikhi A, Barman S and Srivastava G~P 2010 {\em Physical Review B\/} {\bf
  81} 195320

\bibitem{savin2014}
Savin A~V and Kosevich Y~A 2014 {\em Physical Review E\/} {\bf 89} 032102

\bibitem{romensky2014}
Romensky M, Lobaskin V and Ihle T 2014 {\em Phys. Rev. E\/} {\bf 90}(6) 063315

\bibitem{martins2015}
Martins A~C 2015 {\em Physics Letters A\/} {\bf 379} 89 -- 94

\bibitem{andersson1996}
Andersson O and Suga H 1996 {\em Journal of Physics and Chemistry of Solids\/}
  {\bf 57} 125--132

\bibitem{yang2010}
Yang H~S, Cahill D~G, Liu X, Feldman J~L, Crandall R~S, Sperling B~A and
  Abelson J~R 2010 {\em Physical Review B\/} {\bf 81} 104203

\end{thebibliography}


\end{document}